\documentstyle{article}

\begin{document}

\def\levelset{{\cal X}}
\def\fxx{{\bf x}}
\def\fxy{{\bf y}}
\def\real{I\!\!R}

\begin{center}
{\Large Invariability Conditions of Motion Equations of non-Abelian }\\
{\Large Gauge Fields and Elimination of Higgs Mechanism (2) }\\
\vskip 0.2in
{\large Mei Xiaochun}\\
\vskip 0.2in
(Department of Physics, Fuzhou University, Fuzhou, 350025, China, E-mail: mxc001@163.com)\\
\end{center}
\begin{abstract}
\end{abstract}
\par
It is proved that in order to keep both the lagrangian and the motion equation of non-Abelian gauge fields unchanged under the gauge transformation simultaneously, some restriction conditions should be established between the gauge potentials and the group parameters. The result is equivalent to the Faddeev--Popov theory. For non-Abelian gauge fields, it leads to the result that the gauge potentials themselves are unchanged under gauge transformations. In this way, the mass items can be added into the Lagrangian directly without violating gauge invariability and the theory is also renormalizable so that the Higgs mechanism becomes unnecessary.
\\
\\
\par
According to the Yang-Mills theory, in order to keep the Largrangian unchanged under the local gauge transformation, the transformation rules of the field $\phi$ and its covariant differentiation should be defined as
\begin{equation}
\phi'(x)=\exp[-i\theta^{\alpha}(x){T}^{\alpha}]\phi(x)
\end{equation}
\begin{equation}
D'_{\mu}(x)\phi'(x)=\exp[-i\theta^{\alpha}(x){T}^{\alpha}]D_{\mu}(x)\phi(x)
\end{equation}
\begin{equation}
D_{\mu}(x)=\partial_{\mu}+A_{\mu}(x)~~~~~~~~~~~~~A_{\mu}(x)=-igA^{\alpha}_{\mu}(x){T}^{\alpha}
\end{equation}
Here $T^{\alpha}$ are group elements and $\theta^{\alpha}(x)$ are group parameters. The function forms of $\theta^{\alpha}(x)$ are considered arbitrary at present. From Eq.(2), we can get the transformation rule of gauge potentials 
\begin{equation}
A'^{\alpha}_{\mu}(x)=A^{\alpha}_{\mu}(x)+f^{\alpha\beta\gamma}\theta^{\beta}(x){A}^{\gamma}_{\mu}(x)-{1\over{g}}\partial_{\mu}\theta^{\alpha}(x)
\end{equation}
The intensities of gauge fields are defined as
\begin{equation}
F^{\alpha}_{\mu\nu}(x)=\partial_{\mu}{A}^{\alpha}_{\nu}(x)-\partial_{\nu}{A}^{\alpha}_{\mu}(x)+gf^{\alpha\beta\gamma}{A}^{\beta}_{\mu}(x){A}^{\gamma}_{\nu}(x)
\end{equation}
Its transformation rule is
\begin{equation}
F'^{\alpha}_{\mu\nu}(x)=F^{\alpha}_{\mu\nu}(x)+f^{\alpha\beta\gamma}\theta^{\beta}(x){F}^{\gamma}_{\mu\nu}(x)
\end{equation}
The Largrangian of non-Abelian gauge fields with zero masses 
\begin{equation}
L_{0}(x)=-{1\over{4}}{F}^{\alpha}_{\mu\nu}(x){F}^{\alpha}_{\mu\nu}(x)
\end{equation}
is unchanged under gauge transformation. But the Largrangian with mass items can't keep unchanged under gauge transformation. 
\par
However, it is easy to prove that the motion equation of non-Abelian gauge fields can't keep unchanged under gauge transformation in general while the Largrangian of gauge field is invariable. In order to keep the motion equation of gauge field unchanged simultaneously, a restriction relation should be established between the gauge potentials and the group parameters. The free electromagnetic field is discussed at first. Because there exist the Lorentz condition $\partial_{\mu}{A}_{\mu}=0$, free electromagnetic field's motion equation is
\begin{equation}
\partial_{\mu}{F}_{\mu\nu}=\partial^2{A}_{\nu}-\partial_{\nu}\partial_{\mu}{A}_{\mu}=\partial^2{A}_{\nu}=0
\end{equation}
$U(1)$ gauge transformation is defined as:
\begin{equation}
A'_{\nu}(x)=A_{\nu}(x)-{1\over{g}}\partial_{\nu}\theta(x)
\end{equation}
So the gauge transformation of Eq.(8) is
\begin{equation}
\partial_{\mu}{F}'_{\mu\nu}=\partial^2{A}'_{\nu}(x)=\partial^2{A}_{\nu}(x)-{1\over{g}}\partial_{\nu}\partial^2\theta(x)=0
\end{equation}
In order to keep the motion equation unchanged, we must have
\begin{equation}
-\partial_{\nu}\partial^2\theta(x)=0
\end{equation}
It means that $\partial^2\theta=b=constant$, or $\partial_{\nu}\theta=b_{\nu}=constant$, or $\partial_{\nu}\theta=b_{\nu}\neq{constant}$ but $\partial^2{b}_{\nu}=0$. So in order to keep the motion equation of free electromagnetic field unchanged under $U(1)$ gauge transformation, the form of group parameter $\theta(x)$ can't be arbitrary. Only when $\theta(x)$ take three forms above, the motion equation is invariable. As for non-free electromagnetic field, the motion equation is
\begin{equation}
\partial^2{A}_{\mu}(x)=-j_{\mu}~~~~~~~~~~~j_{\mu}=i{e\over{2}}(\bar{\psi}\gamma_{\mu}\psi-\psi^{\tau}\gamma^{\tau}_{\mu}\bar{\psi}^{\tau})
\end{equation}
It is obvious that that the restriction condition Eq.(11) is also effective to keep non-free motion equation unchanged under $U(1)$ transformation. On the other hand, for general $U(1)$ gauge fields, because there exists no Lorentz condition, $\partial_{\mu}{A}_{\mu}\neq{0}$, $U(1)$ gauge transformation of motion equation is
\begin{equation}
\partial_{\mu}{F}'_{\mu\nu}=\partial^2{A}'_{\nu}-\partial_{\nu}\partial_{\mu}{A}'_{\mu}=\partial^2{A}_{\nu}-\partial_{\nu}\partial_{\mu}{A}_{\mu}=\partial_{\mu}{F}_{\mu\nu}
\end{equation}
So the motion equation can keep unchanged without any restriction of group parameter, or $\theta(x)$ can take any function form. 
\par
For $SU(N)$ non-Abelian gauge field, the motion equation is $^{(1)}$ 
\begin{equation}
\partial_{\mu}{F}^{\alpha}_{\mu\nu}+gf^{\alpha\beta\gamma}{A}^{\beta}_{\mu}{F}^{\gamma}_{\mu\nu}=0
\end{equation}
Under the gauge transformation defined in Eq.(4) and (6), the motion equation becomes
$$\partial_{\mu}{F'}^{\alpha}_{\mu\nu}+gf^{\alpha\beta\gamma}{A'}^{\beta}_{\mu}{F'}^{\gamma}_{\mu\nu}=\partial_{\mu}(F^{\alpha}_{\mu\nu}+f^{\alpha\beta\gamma}\theta^{\beta}{F}^{\gamma}_{\mu\nu})$$
\begin{equation}
+gf^{\alpha\beta\gamma}(A^{\beta}_{\mu}+f^{\beta\rho\sigma}\theta^{\rho}{A}^{\sigma}_{\mu}-{1\over{g}}\partial_{\mu}\theta^{\beta})(F^{\gamma}_{\mu\nu}+f^{\gamma\lambda\omega}\theta^{\lambda}{F}^{\omega}_{\mu\nu})=0
\end{equation}
By considering Eq.(14), the formula above becomes
$$f^{\alpha\beta\gamma}[(\partial_{\mu}\theta^{\beta}){F}^{\gamma}_{\mu\nu}+\theta^{\beta}\partial_{\mu}{F}^{\gamma}_{\mu\nu}]+gf^{\alpha\beta\gamma}{f}^{\gamma\lambda\omega}{A}^{\beta}_{\mu}\theta^{\lambda}{F}^{\omega}_{\mu\nu}$$
\begin{equation}
+gf^{\alpha\beta\gamma}(f^{\beta\rho\sigma}\theta^{\rho}{A}^{\sigma}_{\mu}-{1\over{g}}\partial_{\mu}\theta^{\beta})(F^{\gamma}_{\mu\nu}+f^{\gamma\lambda\omega}\theta^{\lambda}{F}^{\omega}_{\mu\nu})=0
\end{equation}
It is easy to see that the solution of the formula above is 
\begin{equation}
\partial_{\mu}\theta^{\alpha}=gf^{\alpha\beta\gamma}\theta^{\beta}{A}^{\gamma}_{\mu}
\end{equation}
In fact, by considering Eq.(14) and (17), as well as the anti-symmetry relation of group construction constant $f^{\alpha\beta\gamma}$, the left side of Eq.(16) becomes
$$f^{\alpha\beta\gamma}{f}^{\beta\rho\sigma}\theta^{\rho}{A}^{\sigma}_{\mu}{F}^{\gamma}_{\mu\nu}-f^{\alpha\beta\gamma}{f}^{\gamma\rho\sigma}\theta^{\beta}{A}^{\rho}_{\mu}{F}^{\sigma}_{\mu\nu}+f^{\alpha\beta\gamma}{f}^{\gamma\rho\sigma}\theta^{\rho}{A}^{\beta}_{\mu}{F}^{\sigma}_{\mu\nu}$$
\begin{equation}
=-(f^{\alpha\sigma\gamma}{f}^{\gamma\rho\beta}+f^{\alpha\rho\gamma}{f}^{\gamma\beta\sigma}+f^{\alpha\beta\gamma}{f}^{\gamma\sigma\rho})\theta^{\rho}{A}^{\beta}_{\mu}{F}^{\sigma}_{\mu\nu}
\end{equation}
By the Jacobian relation $f^{\alpha\sigma\gamma}{f}^{\gamma\rho\beta}+f^{\alpha\rho\gamma}{f}^{\gamma\beta\sigma}+f^{\alpha\beta\gamma}{f}^{\gamma\sigma\rho}=0$, we prove that the Eq.(17) is the solution of Eq.(15). Similarly, for non-free $SU(N)$ non-Abelian gauge field, when there exists interaction between spinor particles and gauge particles, the motion equation of gauge potential is: 
\begin{equation}
\partial_{\mu}{F}^{\alpha}_{\mu\nu}+gf^{\alpha\beta\gamma}{A}^{\beta}_{\mu}{F}^{\gamma}_{\mu}=-ig\bar{\psi}{{\lambda^{\rho}}\over{2}}\gamma_{\mu}\delta_{\mu\nu}\delta_{\alpha\rho}\psi
\end{equation}
It is obvious that that the restriction condition Eq.(17) is also effective to keep the equation unchanged under gauge transformation. It can be known from Eq.(4) that Eq.(17) just means $A'^{\alpha}_{\mu}=A^{\alpha}_{\mu}$, i.e., gauge potential itself is unchanged under gauge transformation. The result notes that if we want to keep the Lagrangian and the motion equation of non-Alberian gauge field invariable simultaneously under gauge transformation, the gauge potentials should be invariable under the transformations. For $SU(N)$ gauge group with $\alpha=1,2\cdot\cdot\cdot{N}$, Eq.(14) represents $4N$ equations. So there are also $4N$ independent group parameters, i.e., there are 4 sets group parameters satisfies Eq.(17). It means that the forms of group parameters are not unique. Because $\theta^{\alpha}(x)\neq{constant}$ in this cases, even though gauge potentials are unchanged under gauge transformation, the local gauge transformations of other fields $\phi(x)$ and their covariant differential defined in Eq.(1) and (2) are still meaningful.
\par
It should be emphasized that motion equation can't contain arbitrary group parameters after gauge transformation, otherwise the motion equation become arbitrary and meaningless. So we call the gauge theory, in which group parameters are arbitrary, as a complete local gauge theory. We call the gauge theory, in which group parameters can't take arbitrary form as incomplete local gauge theory. It can be said that so-called local gauge invariability actually implies incomplete local gauge invariability. For non-Abelian gauge field, the complete local gauge invariability is only a belief without inevitability, for it would destroy the gauge invariability of motion equation in general. It will be proved below that as long as the complete local gauge invariability is replaced by incomplete local gauge invariability, the Higgs mechanics would become unnecessary and the description of gauge theory would also become more symmetrical and simple. 
\par
It is shown below that the result above coincides with the Faddeev-Popov theory $^{(2)}$ In order to avoid infinity, Faddeev and Popov suggested that the integral over function space should be restricted on the hyper-surface decided by gauge condition
\begin{equation}
F[A^{\alpha}_{\mu}]=0~~~~~~~~~~~~~~~~\alpha=1,2,\cdot\cdot\cdot{N}
\end{equation}
In this way, the freedom degrees of gauge fields are decreased from $4N$ to $3N$. The following relation is used to restrict orbit integral
\begin{equation}
\triangle_F[A^{\alpha}_{\mu}]\cdot\int[dg]\delta(F[A^{\alpha{g}}_{\mu}])=1
\end{equation}
The restriction condition $\delta(F\mid{A}^{\alpha{g}}_{\mu}\mid)$ demands $F\mid{A}^{\alpha{g}}_{\mu}\mid=0$ actually. For $U(1)$ gauge field, by taking the Landau gauge condition $F\mid{A}^{\alpha}_{\mu}\mid=\partial_{\mu}{A}_{\mu}=0$, we have
\begin{equation}
F[A^{g}_{\mu}]=0\rightarrow\partial_{\mu}{A}'_{\mu}=\partial_{\mu}(A_{\mu}-{1\over{g}}\partial_{\mu}\theta)=-{1\over{g}}\partial^2\theta=0
\end{equation}
The result coincides with Eq.(11) (to take $b=0$). For non-Abelian gauge fields, by the Landau gauge condition $\partial_{\mu}{A}^{\alpha}_{\mu}=0$, we have
\begin{equation}
F[A^{\alpha{g}}_{\mu}]=0\rightarrow\partial_{\mu}{A'}^{\alpha}_{\mu}=\partial_{\mu}(A^{\alpha}_{\mu}+f^{\alpha\beta\gamma}\theta^{\beta}{A}^{\gamma}_{\mu}-{1\over{g}}\partial_{\mu}\theta^{\alpha})=\partial_{\mu}(f^{\alpha\beta\gamma}\theta^{\beta}{A}^{\gamma}_{\mu}-{1\over{g}}\partial_{\mu}\theta^{\alpha})=0
\end{equation}
The formula can be written as 
\begin{equation}
f^{\alpha\beta\gamma}\theta^{\beta}{A}^{\gamma}_{\mu}-{1\over{g}}\partial_{\mu}\theta^{\alpha}=b_{\mu}~~~~~~~~~~~~~~~~~~\partial_{\mu}{b}_{\mu}=0
\end{equation}
Taking the simplest form to let $b_{\mu}=0$, we get Eq.(17). So the result in this paper coincides with the Faddeev-Popov theory, or it is the simplest form of the Faddeev-Popov theory. It can be seen that though restriction relation (23) can eliminate the infinity of orbit integral, it can't make motion equation invariable under gauge transformation. To make both the orbit integral and the motion equation of non-Abelian gauge fields invariable, the restriction relation $\delta(\partial_{\mu}{b}_{\mu})$ should be changed into $\delta(b_{\mu})$. So Eq.(21) should be re-written as 
\begin{equation}
\triangle_{F}[A^{\alpha}_{\mu}]\cdot\int[dg]\delta[R_{\mu}(f^{\alpha\beta\gamma}\theta^{\beta}{A}^{\gamma}_{\mu}-{1\over{g}}\partial_{\mu}\theta^{\alpha})]=1
\end{equation}
Here $R_{\mu}$ is an arbitrary constant vector so that Eq.(17) is always tenable. In this way, by the relation $\triangle_{F}[A^{\alpha}_{\mu}]=detM_{F}$  we have
\begin{equation}
M^{\alpha\beta}_{F}(x,y)=R_{\mu}[f^{\alpha\beta\gamma}{A}^{\gamma}_{\mu}(y)\delta^{4}(x-y)-{1\over{g}}\delta_{\alpha\beta}\partial_{\mu}\delta^4(x-y)]
\end{equation}
The action of ghost particle becomes
\begin{equation}
S_g=\int{d}^4{x}{C}^{+}_{\alpha}{R}_{\mu}(\delta_{\alpha\beta}\partial_{\mu}-gf^{\alpha\beta\gamma}{A}^{\gamma}_{\mu}){C}_{\beta}
\end{equation}
It can be seen below that the change of ghost particle's action has no essential influence on interaction theory for ghost particles are fictitious.
\par
Because gauge potential is unchanged under gauge transformation with $A'^{\alpha}_{\mu}=A^{\alpha}_{\mu}$ for non-Abelian gauge field, we can add mass item into the Lagrangian directly without violating gauge invariability. Therefore, the Lagrangian below is invariable under gauge transformation
\begin{equation}
L=-{1\over{4}}{F}^{\alpha}_{\mu\nu}{F}^{\alpha}_{\mu\nu}-{1\over{2}}{m}^2_{A}{A}^{\alpha}_{\mu}{A}^{\alpha}_{\mu}
\end{equation}
It can be proved that when the interaction between gauge particles and other particles are considered, corresponding $W,T$ identity can also be obtained and the theory is also renormalizable. For simplification at first, we consider a system composed of gauge field $A^{\alpha}_{\mu}$, Fermi field $\psi$, and ghost fields $C^{+}_{\alpha}$ and $C_{\alpha}$. Let $S_f$ represent the action of gauge and Fermi fields, $S_h$ represent the action of fixed gauge item, $S_g$ represent the action of ghost field, after mass item is added, the effective action of the system is $S_{eff}=S_f+S_h+S_g$ with
$$S_f=\int{d}^4{x}[-\bar{\psi}(\gamma_{\mu}\partial_{\mu}-ig{{\tau_{\alpha}}\over{2}}\gamma_{\mu}{A}^{\alpha}_{\mu}+m_{\psi})\psi$$
\begin{equation}
-{1\over{4}}(\partial_{\mu}{A}^{\alpha}_{\nu}-\partial_{\nu}{A}^{\alpha}_{\mu}+gf^{\alpha\beta\gamma}{A}^{\beta}_{\mu}{A}^{\gamma}_{\nu})^2-{1\over{2}}{m}_A{A}^{\alpha}_{\mu}{A}^{\alpha}_{\mu}]
\end{equation}
\begin{equation}
S_h=\int{d}^4{x}[-{1\over{2\zeta}}(\partial_{\mu}{A}^{\alpha}_{\mu})^2]~~~~~~~~~~~~~~S_g=\int{d}^4{x}{C}^{+}_{\alpha}{R}_{\mu}(\delta_{\alpha\beta}\partial_{\mu}-gf^{\alpha\beta\gamma}{A}^{\gamma}_{\mu}){C}_{\beta}
\end{equation}
Because $A^{\alpha}_{\mu}$ is invariable according to this paper, $S_f$ and $S_h$ are invariable under $SU(N)$ gauge transformation. Because $\triangle_{F}[A^{\alpha}_{\mu}]$ is unchanged under gauge transformation, so ghost fields $C^{+}_{\alpha}$ and $C_{\alpha}$can also be regarded invariable under gauge transformation, though in the current they are not. In this way, the simplified $B,R,S$ transformations can be written as
\begin{equation}
\delta\bar{\psi}=-i\bar{\psi}{{\tau_{\alpha}}\over{2}}{C}_{\alpha}\delta\lambda~~~~~~~~~~~~~~~~~~~\delta\psi=i{{\tau_{\alpha}}\over{2}}{C}_{\alpha}\psi\delta\lambda
\end{equation}
\begin{equation}
\delta{A}^{\alpha}_{\mu}=0~~~~~~~~~~~~~~\delta{C}^{+}_{\alpha}=0~~~~~~~~~~~~~~~\delta{C}_{\alpha}=0
\end{equation}
Here $\delta\lambda$ is infinitesimal with $(\delta\lambda)^{2}\rightarrow{0}$. Similarly, we also have $\delta^2\bar{\psi}=0$ and $\delta^{2}\psi=0$. The similar generating function of the Green's function, that is unchanged under simplified transformations (31) and (32), can be written as 
$$Z=\int[d\bar{\psi}][d\psi][dA^{\alpha}_{\mu}][dC^{+}_{\alpha}][dC_{\alpha}]\exp\{iS_{eff}$$
\begin{equation}
+i\int{d}x^4[\bar{\eta}\psi+\bar{\psi}\eta+J^{\alpha}_{\mu}{A}^{\alpha}_{\mu}+\varsigma_{\alpha}{C}^{+}_{\alpha}+\varsigma^{+}_{\alpha}{C}_{\alpha}+\bar{K}\delta\psi+\delta\bar{\psi}{K}]\}
\end{equation}
Here $K$ and $\bar{K}$ are anti-commutative quantities. Because integral is independent of variable transformations, $S_{eff}$ is invariable under gauge transformation, as well as $\delta^2\bar{\psi}=\delta^2\psi=0$, the formula above is unchanged under transformations $\bar{\psi}\rightarrow\bar{\psi'}=\bar{\psi}+\delta\bar{\psi}$, $\psi\rightarrow\psi'=\psi+\delta\psi$. So we can also write it as 
$$Z=\int[d\bar{\psi}][d\psi][dA^{\alpha}_{\mu}][dC^{+}_{\alpha}][dC_{\alpha}]\exp\{iS_{eff}$$
\begin{equation}
+i\int{d}x^4[\bar{\eta}(\psi+\delta\psi)+(\bar{\psi}+\delta\bar{\psi})\eta+J^{\alpha}_{\mu}{A}^{\alpha}_{\mu}+\varsigma_{\alpha}{C}^{+}_{\alpha}+\varsigma^{+}_{\alpha}{C}_{\alpha}+\bar{K}\delta\psi+\delta\bar{\psi}{K}]\}
\end{equation}
Eq.(34) minus Eq.(33), we get
$$\int[d\bar{\psi}][d\psi][dA^{\alpha}_{\mu}][dC^{+}_{\alpha}][dC_{\alpha}]\int{d}x^4[\bar{\eta}\delta\psi+\delta\bar{\psi}\eta]$$
\begin{equation}
\times\exp[iS_{eff}+i\int{d}x^4(\bar{\eta}\psi+\bar{\psi}\eta+J^{\alpha}_{\mu}{A}^{\alpha}_{\mu}+\varsigma_{\alpha}{C}^{+}_{\alpha}+\varsigma^{+}_{\alpha}{C}_{\alpha}+\bar{K}\delta\psi+\delta\bar{\psi}{K})]=0
\end{equation}
Let $\delta\psi\rightarrow\delta/(i\delta\bar{K})$, $\delta\bar{\psi}\rightarrow\delta/(i\delta{K})$, the simplified $W,T$ identity represented by the generating function of the Green's function can be written as
\begin{equation}
[\bar{\eta}{{\delta}\over{\delta\bar{K}}}+{{\delta}\over{\delta{K}}}\eta]Z(\bar{\eta},\eta,J^{\alpha}_{\mu},\varsigma^{+},\varsigma,\bar{K},K)=0
\end{equation}
The simplified $W,T$ identity represented by the generating function of normal vertex angle becomes
\begin{equation}
{{\delta\Gamma}\over{\delta\psi}}{{\delta\Gamma}\over{\delta\bar{K}}}+{{\delta\Gamma}\over{\delta{K}}}{{\delta\Gamma}\over{\delta\bar{\psi}}}=\Gamma\ast\Gamma=0
\end{equation}
But there is no ghost equation for $SU(N)$ gauge fields. The normalization of single loop approximation is discussed below. It is only a simplified process of current theory. After items $\bar{K}\delta\psi$ and $\delta\bar{\psi}{K}$ are added into the action, the effective action unchanged under simplified $B,R,S$ transformations can be written as
$$S_0=\int{d}^4{x}[-\bar{\psi}(\gamma_{\mu}\partial_{\mu}-ig{{\tau_{\alpha}}\over{2}}\gamma_{\mu}{A}^{\alpha}_{\mu}+m_{\psi})\psi$$
$$-{1\over{4}}(\partial_{\mu}{A}^{\alpha}_{\nu}-\partial_{\nu}{A}^{\alpha}_{\mu}+gf^{\alpha\beta\gamma}{A}^{\beta}_{\mu}{A}^{\gamma}_{\nu})^2-{1\over{2}}{m}^2_{A}{A}^{\alpha}_{\mu}{A}^{\alpha}_{\mu}-{1\over{2\zeta}}(\partial_{\mu}{A}^{\alpha}_{\mu})^2$$
\begin{equation}
+C^{+}_{\alpha}{R}_{\mu}(\delta_{\alpha\beta}\partial_{\mu}-gf^{\alpha\beta\gamma}{A}^{\gamma}_{\mu}){C}_{\beta}+ig\bar{K}{{\tau_{\alpha}}\over{2}}{C}_{\alpha}\psi\delta\lambda-ig\bar{\psi}{{\tau_{\alpha}}\over{2}}{C}_{\alpha}{K}\delta\lambda]
\end{equation}
Using it to construct the generating function of normal vertex angle, we obtain $\Gamma[S_0]\simeq{S}_0$ under tree approximation. The process is finite. Therefore, according to Eq.(37), we have $S_0\ast{S}_0=0$. For single loop approximation, we can write
\begin{equation}
\Gamma[S_0]=S_0+\Gamma^{f}_1[S_0]+\Gamma^{d}_1[S_0]
\end{equation}
Here $\Gamma^{f}_1[S_0]$ is finite but $\Gamma^{d}_1[S_0]$ is infinite. In order to eliminate infinite, for single loop approximation, we use $S_{0}+\triangle{S}_0$ to construct the generating function of normal vertex angle 
\begin{equation}
\Gamma[S_0+\triangle{S}_0]\simeq{S}_0+\triangle{S}_0+\Gamma^{f}_1[S_0]+\Gamma^{d}_1[S_0]
\end{equation}
By taking $\triangle{S}_0=-\Gamma^{f}_1[S_0]$, the infinite of single loop approximation can be eliminated It can be proved below that we also have
\begin{equation}
-\Gamma^{d}_1[S_0]=\sum_{\sigma}{a}_{\sigma}{G}_{\sigma}+\hat{S}_0\ast{F}~~~~~~~~~~~~~~~~S_{0}\ast\Gamma^{d}_{1}[S_0]=0
\end{equation}
Here $G_{\sigma}$ is an invariable quantity of gauge transformation with form 
$$\sum_{\sigma}{a}_{\sigma}{G}_{\sigma}=\int{d}^4{x}[-a_1\bar{\psi}(\gamma_{\mu}\partial_{\mu}-ig{{\lambda_{\alpha}}\over{2}}\gamma_{\mu}{A}^{\alpha}_{\mu})\psi-a_2{m}_{\psi}\bar{\psi}\psi$$
$$-{1\over{4}}{a}_3(\partial_{\mu}{A}^{\alpha}_{\nu}-\partial_{\nu}{A}^{\alpha}_{\mu}+gf^{\alpha\beta\gamma}{A}^{\beta}_{\mu}{A}^{\gamma}_{\nu})^2-{1\over{2}}{a}_4{m}^2_{A}{A}^{\alpha}_{\mu}{A}^{\alpha}_{\mu}-{1\over{2\zeta}}{a}_5(\partial_{\mu}{A}^{\alpha}_{\mu})^2$$
\begin{equation}
+a_6{C}^{+}_{\alpha}{R}_{\mu}(\delta_{\alpha\beta}\partial_{\mu}-gf^{\alpha\beta\gamma}{A}^{\gamma}_{\mu}){C}_{\beta}]
\end{equation}
Here $a_i$ are constants containing infinite polos. Because there is no ghost equation, $F$ can be an arbitrary function. We can also write it as similarly
\begin{equation}
F=\int{d}^4{x}(b_1\bar{K}\psi+b_2\bar{\psi}K)
\end{equation}
Because $G_{\sigma}$ does not contain $K$ and $\bar{K}$, according to Eq.(33), we have $S_0\sim\bar{K}\delta\psi+\delta\bar{\psi}{K}$, so
$$S_0\ast{G}_{\sigma}={{\delta{S}_0}\over{\delta\psi}}{{\delta{G}_{\sigma}}\over{\delta\bar{K}}}+{{\delta{S}_0}\over{\delta\bar{K}}}{{\delta{G}_{\sigma}}\over{\delta\psi}}+{{\delta{S}_0}\over{\delta\bar{\psi}}}{{\delta{G}_{\sigma}}\over{\delta{K}}}+{{\delta{S}_0}\over{\delta{k}}}{{\delta{G}_{\sigma}}\over{\delta\bar{\psi}}}$$
\begin{equation}
={{\delta{G}_{\sigma}}\over{\delta\bar{\psi}}}\delta\bar{\psi}+{{\delta{G}_{\sigma}}\over{\delta\psi}}\delta\psi=\delta{G}_{\sigma}=0
\end{equation}
By the anti-commutation relation between $\psi$ and $\bar{\psi}$, it can also be proved as done in the current theory
\begin{equation}
S_0\ast(S_0\ast{F})=0
\end{equation}
So $\Gamma^{d}_1[S_0]$ satisfies Eq.(41) and we obtain
$$S_0+\triangle{S}_0=S_0+\sum_{\sigma}{a}_{\sigma}{G}_{\sigma}+\hat{S}_{0}\ast{F}$$
\begin{equation}
+S_0+{{\delta{S}_0}\over{\delta\psi}}{{\delta{F}}\over{\delta\bar{K}}}+{{\delta{S}_0}\over{\delta\bar{K}}}{{\delta{F}}\over{\delta\psi}}+{{\delta{S}_0}\over{\delta\bar{\psi}}}{{\delta{F}}\over{\delta{K}}}+{{\delta{S}_0}\over{\delta{k}}}{{\delta{F}}\over{\delta\bar{\psi}}}+\sum_{\sigma}{a}_{\sigma}{G}_{\sigma}
\end{equation}
On the other hand, according to the current theory, let
\begin{equation}
\bar{\psi}'=\bar{\psi}+{{\delta{F}}\over{\delta\bar{K}}}=(1+b_1\bar{\psi})=Y_{\bar{\psi}}\bar{\psi}~~~~~~~~~~\psi'=\psi+{{\delta{F}}\over{\delta{K}}}=(1+b_2\psi)=Y_{\psi}\bar{\psi}
\end{equation}
\begin{equation}
K'=K-{{\delta{F}}\over{\delta\bar{\psi}}}=(1-b_1\bar{K})=Y^{-1}_{\bar{\psi}}{K}~~~~~~~~~~\bar{K}'=\bar{K}-{{\delta{F}}\over{\delta\psi}}=(1-b_2\bar{K})=Y^{-1}_{\psi}\bar{K}
\end{equation}
it can be proved $^{(3)}$£º
\begin{equation}
S_{0}(\bar{\psi}',\psi',\bar{K}',K')=S_0(\bar{\psi},\psi,\bar{K},K)+{{\delta{S}_0}\over{\delta\psi}}{{\delta{F}}\over{\delta\bar{K}}}+{{\delta{S}_0}\over{\delta\bar{K}}}{{\delta{F}}\over{\delta\psi}}+{{\delta{S}_0}\over{\delta\bar{\psi}}}{{\delta{F}}\over{\delta{K}}}+{{\delta{S}_0}\over{\delta{k}}}{{\delta{F}}\over{\delta\bar{\psi}}}
\end{equation}
Put the formula above into Eq. (46), it can be known that the effect of item $S_0\ast{F}$ is to do the transformations of Eq.(46) and (47) in the action $S_0$. So we can define function $G_{\sigma}$ by using $\bar{\psi}',\psi',\bar{K}',K'$ at beginning. Then let $Y_1=1+a_1$, $Y_3=1+a_2$, $Y_3=1+a_3$, $Y_4=1+a_4$, $Y_5=1+a_5$, $Y_6=1+a_6$. In this way, the action of renormalization in single loop process can be written as
$$S_1=S_0+\triangle{S}_0=\int{d}^4{x}[-Y_{\bar{\psi}}{Y}_{\psi}\bar{\psi}(\gamma_{\mu}\partial_{\mu}-ig{{\tau_{\alpha}}\over{2}}\gamma_{\mu}{A}^{\alpha}_{\mu})\psi-Y_{\bar{\psi}}{Y}_{\psi}{m}_{\psi}\bar{\psi}\psi$$
$$-{1\over{4}}(\partial_{\mu}{A}^{\alpha}_{\nu}-\partial_{\nu}{A}^{\alpha}_{\mu}+gf^{\alpha\beta\gamma}{A}^{\beta}_{\mu}{A}^{\gamma}_{\nu})^2-{1\over{2}}{m}^2_{A}{A}^{\alpha}_{\mu}{A}^{\alpha}_{\mu}-{1\over{2\zeta}}(\partial_{\mu}{A}^{\alpha}_{\mu})^2$$
$$+C^{+}_{\alpha}{R}_{\mu}\partial_{\mu}{C}_{\alpha}-gf^{\alpha\beta\gamma}{C}^{+}_{\alpha}{R}_{\mu}{A}^{\beta}_{\mu}{C}_{\gamma}+ig\bar{K}{{\tau_{\alpha}}\over{2}}{C}_{\alpha}\psi\delta\lambda-ig\bar{\psi}{{\tau_{\alpha}}\over{2}}{C}_{\alpha}{K}\delta\lambda$$
$$-(Y_1-1)Y_{\bar{\psi}}{Y}_{\psi}\bar{\psi}(\gamma_{\mu}\partial_{\mu}-ig{{\tau_{\alpha}}\over{2}}\gamma_{\mu}{A}^{\alpha}_{\mu})-(Y_2-1)Y_{\bar{\psi}}{Y}_{\psi}{m}_{\psi}\bar{\psi}\psi$$
$$-{1\over{4}}(Y_3-1)(\partial_{\mu}{A}^{\alpha}_{\nu}-\partial_{\nu}{A}^{\alpha}_{\mu}+gf^{\alpha\beta\gamma}{A}^{\beta}_{\mu}{A}^{\gamma}_{\nu})^2-{1\over{2}}(Y_4-1)m^2_{A}{A}^{\alpha}_{\mu}{A}^{\alpha}_{\mu}$$
$$-{1\over{2\zeta}}(Y_5-1)(\partial_{\mu}{A}^{\alpha}_{\nu})^2+(Y_6-1)(C^{+}_{\alpha}{R}_{\mu}\partial_{\mu}{C}_{\alpha}-gf^{\alpha\beta\gamma}{C}^{+}_{\alpha}{R}_{\mu}{A}^{\beta}_{\mu}{C}_{\gamma})]$$
$$=\int{d}^4{x}[-Y_1{Y}_{\bar{\psi}}{Y}_{\psi}\bar{\psi}(\gamma_{\mu}\partial_{\mu}-ig{{\tau_{\alpha}}\over{2}}\gamma_{\mu}{A}^{\alpha}_{\mu})\psi-Y_2{Y}_{\bar{\psi}}{Y}_{\psi}{m}_{\psi}\bar{\psi}\psi$$
$$-{1\over{4}}{Y}_3(\partial_{\mu}{A}^{\alpha}_{\nu}-\partial_{\nu}{A}^{\alpha}_{\mu})^2-{1\over{2}}{Y}_3{g}(\partial_{\mu}{A}^{\alpha}_{\nu}-\partial_{\nu}{A}^{\alpha}_{\mu})f^{\alpha\beta\gamma}{A}^{\beta}_{\mu}{A}^{\gamma}_{\nu}$$
$$-{1\over{4}}{Y}_3(gf^{\alpha\beta\gamma}{A}^{\beta}_{\mu}{A}^{\gamma}_{\nu})^2-{1\over{2}}{Y}_4{m}^2_{A}{A}^{\alpha}_{\mu}{A}^{\alpha}_{\mu}-{1\over{2\zeta}}{Y}_5(\partial_{\mu}{A}^{\alpha}_{\mu})^2$$
\begin{equation}
+Y_6(C^{+}_{\alpha}{R}_{\mu}\partial_{\mu}{C}_{\alpha}-gf^{\alpha\beta\gamma}{C}^{+}_{\alpha}{R}_{\mu}{A}^{\beta}_{\mu}{C}_{\gamma})+ig\bar{K}{{\tau_{\alpha}}\over{2}}{C}_{\alpha}\psi\delta\lambda-ig\bar{\psi}{{\tau_{\alpha}}\over{2}}{C}_{\alpha}{K}\delta\lambda]
\end{equation}
On the other hand, when the action is represented by nude quantities, we have
$$S_1=\int{d}^4{x}[-\bar{\psi}_0(\gamma_{\mu}\partial_{\mu}-ig_0{{\tau_{\alpha}}\over{2}}\gamma_{\mu}{A}^{\alpha}_{0\mu}+m_{0\psi})\psi_0$$
$$-{1\over{4}}(\partial_{\mu}{A}^{\alpha}_{0\nu}-\partial_{\nu}{A}^{\alpha}_{0\mu}+g_0{f}^{\alpha\beta\gamma}{A}^{\beta}_{0\mu}{A}^{\gamma}_{0\nu})^2-{1\over{2}}{m}^2_{0A}{A}^{\alpha}_{0\mu}{A}^{\alpha}_{0\mu}-{1\over{2\zeta_0}}(\partial_{\mu}{A}^{\alpha}_{0\mu})^2$$
\begin{equation}
+C^{+}_{0\alpha}{R}_{\mu}\partial_{\mu}{C}_{0\alpha}-g_0{f}^{\alpha\beta\gamma}{C}^{+}_{0\alpha}{R}_{\mu}{A}^{\beta}_{0\mu}{C}_{0\gamma}+ig_{0}\bar{K}_0{{\tau_{\alpha}}\over{2}}{C}_{0\alpha}\psi_0\delta\lambda-ig_0\bar{\psi}_0{{\tau_{\alpha}}\over{2}}{C}_{0\alpha}{K}_0\delta\lambda]
\end{equation}
Let $\psi_0=\sqrt{Z_2}\psi$, $\bar{\psi}_0=\sqrt{Z_2}\bar{\psi}$, $A^{\alpha}_{0\mu}=\sqrt{Z_3}{A}^{\alpha}_{\mu}$, $C^{+}_{0\alpha}=\sqrt{\tilde{Z}_3}{C}^{+}_{\alpha}$, $C_{0\alpha}=\sqrt{\tilde{Z}_3}{C}_{\alpha}$, $\bar{K}_0=\sqrt{Z_K}\bar{K}$, $K_0=\sqrt{Z_K}{K}$, $g_0=Z_{gi}{g}$, $m_{0\psi}=Z_{m\psi}{m}_{\psi}$, $M_{0A}=Z_{mA}{m}_A$, $\varsigma_0=Z_{\varsigma}\varsigma$, the formula above becomes
$$S_1=\int{d}^4{x}[-Z_2\bar{\psi}(\gamma_{\mu}\partial_{\mu}-iZ_{g1}\sqrt{Z_3}{g}{{\tau_{\alpha}}\over{2}}\gamma_{\mu}{A}^{\alpha}_{\mu}+Z_{m\psi}{m}_{\psi})\psi$$
$$-{1\over{4}}{Z}_3(\partial_{\mu}{A}^{\alpha}_{\nu}-\partial_{\nu}{A}^{\alpha}_{\mu})^2-{1\over{2}}{Z}^{3/2}_{3}{Z}_{g2}{g}(\partial_{\mu}{A}^{\alpha}_{\nu}-\partial_{\nu}{A}^{\alpha}_{\mu})f^{\alpha\beta\gamma}{A}^{\beta}_{\mu}{A}^{\gamma}_{\nu}$$
$$-{1\over{4}}{Z}^{2}_{g3}{Z}^{2}_{3}(gf^{\alpha\beta\gamma}{A}^{\beta}_{\mu}{A}^{\gamma}_{\nu})^2-{1\over{2}}{Z}^{2}_{mA}{Z}_{3}{m}^{2}_{A}{A}^{\alpha}_{\mu}{A}^{\alpha}_{\mu}-{1\over{2\zeta}}{Z}_{\zeta}{Z}_3(\partial_{\mu}{A}^{\alpha}_{\mu})^2$$
$$+\tilde{Z}_3{R}_{\mu}{C}^{+}_{\alpha}\partial_{\mu}{C}_{\alpha}-Z_{g4}\tilde{Z}_3\sqrt{Z_3}{g}f^{\alpha\beta\gamma}{C}^{+}_{\alpha}{R}_{\mu}{A}^{\beta}_{\mu}{C}_{\gamma}$$
\begin{equation}
+iZ_{g5}{Z}_{K}\sqrt{\tilde{Z_3}{Z_2}}{g}\bar{K}{{\tau_{\alpha}}\over{2}}{C}_{\alpha}\psi\delta\lambda-iZ_{g6}{Z}_{K}\sqrt{\tilde{Z_3}{Z_2}}{g}\bar{\psi}{{\tau_{\alpha}}\over{2}}{C}_{\alpha}{K}\delta\lambda]
\end{equation}
Comparing the corresponding items between Eq.(50) and (52), we get
$$Z_2=Y_1{Y}_{\bar{\psi}}{Y}_{\psi}~~~~~~~~~~Z_{g1}\sqrt{Z_3}=1~~~~~~~~~Z_2{Z}_{m\psi}=Y_2{Y}_{\bar{\psi}}{Y}_{\psi}~~~~~~~~~~Z_3=Y_3$$
$$Z_{g2}\sqrt{Z_3}=1~~~~~~~~~~Z^{2}_{g3}{Z}_3=1~~~~~~~~~~Z^{2}_{mA}{Z}_3=Y_4~~~~~~~~~~Z_{\zeta}{Z}_3=Y_5$$
\begin{equation}
\tilde{Z}_3=Y_6~~~~~~~~~~~Z_{g4}\sqrt{Z_3}=1~~~~~~~~~~~Z_{g5}{Z}_{K}\sqrt{\tilde{Z}_3{Z}_2}=1~~~~~~~~~~Z_{g6}=Z_{g5}
\end{equation}
It can be obtained immediately
\begin{equation}
Z_{g1}=Z_{g2}=Z_{g3}=Z_{g4}=Z_{g}={1\over{\sqrt{Z_3}}}={1\over{\sqrt{Y_3}}}
\end{equation}
By taking $Z_K=Y_3/\sqrt{\tilde{Z_3}{Z_2}}$ (similar to the current theory), we have $Z_{g5}=Z_{g6}=Z_{g}$. Therefore, the renormalization interaction constants in all items are the same so that renormalization is possible. So for the process of single loop approximation, according to the paper, renormalization constants are taken as
$$Z_2=Y_{1}{Y}_{\bar{\psi}}{Y}_{\psi}~~~~~~~~~~~Z_3=Y_3~~~~~~~~~~~\tilde{Z}_3=Y_6~~~~~~~~~Z_g={1\over{\sqrt{Y_3}}}$$
\begin{equation}
Z_{m\psi}={{Y_2}\over{Y_1}}~~~~~~~~~~~Z_{mA}=\sqrt{{Y_4}\over{Y_3}}~~~~~~~~~Z_{\zeta}={{Y_5}\over{Y_3}}~~~~~~~~~Z_K={{\sqrt{Y_3}}\over{\sqrt{Y_6{Y}_1{Y}_{\bar{\psi}}{Y}_{\psi}}}}
\end{equation}
By the way, because there is no restriction of ghost equation for $SU(N)$ fields, the function $F$ in Eq.(41) can be arbitrary. For simplification, we can take $F=0$ directly so that it is unnecessary for us to introduce Eq.(47) and (48) agin. In this case we have $Y_{\bar{\psi}}=Y_{\psi}=1$ in (55).For higher order processes, renormalization can also be carried out by the similar procedure in the current theory.
\par
The mass item's gauge transformation in the united theory of weak-electric interaction is discussed at last. We only discuss the transformation of lepton field's mass items. The result is suitable to quark fields. In the united theory, we use chiral fields to describe weak interaction. The transformation rules of left hand and right hand fields under $SU(2)\times{U}(1)$ gauge transformation are
\begin{equation}
L\rightarrow{L}'=\exp(-i{{\vec{\theta'}\cdot\vec{\tau}}\over{2}}+i{{\theta}\over{2}})L~~~~~~~~~~~~~~~L=\mid^{\nu_{L}}_{l_{L}}\mid
\end{equation}
\begin{equation}
l_{R}\rightarrow\exp(i\theta){l}_{R}~~~~~~~~~~l_{L}={1\over{2}}(1+\gamma_5)l~~~~~~~~~l_R={1\over{2}}(1-\gamma_5)l
\end{equation}
The Lagrangian of free lepton field without mass item is 
\begin{equation}
L_0=-\bar{L}\gamma_{\mu}\partial_{\mu}{L}-\bar{l}_{R}\gamma_{\mu}\partial_{\mu}{l}_{R}=-\nu_{L}\gamma_{\mu}\partial_{\mu}\nu_{L}-\bar{l}\gamma_{\mu}\partial_{\mu}{l}
\end{equation}
Because the transformation rule of left hand field is different from right hand field, the mass item of lepton field with form
\begin{equation}
m_{l}\bar{l}{l}=m_{l}(\bar{l}_{L}{l}_{R}+\bar{l}_{R}{l}_{L})
\end{equation}
can't not keep unchanged under $SU(2)\times{U}(1)$ transformation. Similar to gauge field's mass item, in the current theory, the mass items of lepton fields can't yet be added into the Lagrangian directly. The Higgs mechanics is needed. It is proved below that this problem can be resolved solved, by establishing a proper restriction relation between group parameters. The Higgs mechanics. According to Eq.(54), we have infinite transformations
\begin{equation}
\nu'_{L}\simeq[1-{i\over{2}}(\theta_3-\theta)]\nu_{L}-{i\over{2}}(\theta_1-i\theta_2){l}_{L}~~~~~~~\nu'\simeq[1-{i\over{2}}(\theta_3-\theta)]\nu-{i\over{2}}(\theta_1-i\theta_2)l
\end{equation}
\begin{equation}
l'_{L}\simeq-{i\over{2}}(\theta_1+i\theta_2)\nu_{L}+[1-{i\over{2}}(\theta_3+\theta)]{l}_{L}~~~~~~~l'\simeq-{i\over{2}}(\theta_1+i\theta_2)\nu+[1-{i\over{2}}(\theta_3+\theta)]l
\end{equation}
If choosing $\theta_1=-i\theta_2$, we get
\begin{equation}
l'\simeq[1-{i\over{2}}(\theta_3+\theta)]l\simeq\exp[-{i\over{2}}(\theta_3+\theta)]l
\end{equation}
In this case, we have $\bar{l}'{l}'=\bar{l}{l}$, the lepton mass item can keep unchanged under $SU(2)\times{U}(1)$ transformation and can be added into the Lagrangian directly. 
\par
For the transformation of gauge field's mass items, the relations between mass eigen states and non-mass eigen states of gauge particles are  
\begin{equation}
W^{+}_{\mu}={1\over{\sqrt{2}}}(A^{1}_{\mu}+iA^{2}_{\mu})~~~~~~~~~~~~W^{-}_{\mu}={1\over{\sqrt{2}}}(A^{1}_{\mu}-iA^{2}_{\mu})
\end{equation}
\begin{equation}
Z_{\mu}=\cos\vartheta_{w}{A}^3_{\mu}-\sin\vartheta_{w}{B}_{\mu}~~~~~~~~~A_{\mu}=\sin\vartheta_{w}{A}^3_{\mu}+\cos\vartheta_{w}{B}_{\mu}
\end{equation}
Here $\theta_{w}$ is the Weiberge angle, $A_{\mu}$ is electromagnetic field. Because $B_{\mu}$ field is massless, by using the formulas above, when the mass items are represented by both mass eigen states and non-mass eigen states, we have relation
\begin{equation}
-{1\over{2}}{m}^2_{A}{A}^{\alpha}_{\mu}{A}^{\alpha}_{\mu}=-m^2_{A}{W}^{+}_{\mu}{W}^{-}_{\mu}+{1\over{2}}{m}^2_{A}\cos\vartheta^2_{w}{Z}_{\mu}{Z}_{\mu}-Q
\end{equation}
\begin{equation}
Q={1\over{2}}{m}^2_{A}\sin\vartheta_{w}(A_{\mu})^2+m^{2}_{A}\sin\vartheta_{w}\cos\vartheta_{w}{A}_{\mu}{Z}_{\mu}
\end{equation}
In the formula, $m_{A}\sin\theta_w$ is photon's mass and product item $A_{\mu}{Z}_{\mu}$ represent two point's interaction. Because theses two items do not exist actually, we should cancel them in the action. As taking $R_{\zeta}$ gauge in the current theory, we take gauge
\begin{equation}
F^{\alpha}[A^{\alpha}_{\mu}]=\partial_{\mu}{A}^{\alpha}_{\mu}+R^{\alpha}~~~~~~~~~~~~R^{\alpha}=-\partial_{\mu}{A}^{\alpha}_{\mu}\pm\sqrt{1-2\zeta_{A}{Q}/\partial_{\mu}{A}^{\alpha}_{\mu}}
\end{equation}
So the gauge fixed item can be written as
\begin{equation}
S_h=\int{d}^4{x}[-{1\over{2\zeta_{A}}}(\partial_{\mu}{A}^{\alpha}_{\mu})^2+Q]
\end{equation}
The superfluous $Q$ in the action produced by Eq.(65) can be canceled. Let $m_A=m_w$, according to (65), we have
\begin{equation}
{1\over{2}}m^2_{A}{A}^{\alpha}_{\mu}{A}^{\alpha}_{\mu}\sim{m}^2_{W}{W}^{+}_{\mu}{W}^{-}_{\mu}+{1\over{2}}m^2_{w}\cos\vartheta^2_{w}{Z}_{\mu}{Z}_{\mu}
\end{equation}
Because $m_{w}\cos\theta_w$ is $Z^{0}$ particle's mass actually, we have
\begin{equation}
m_Z=m_w\cos\vartheta_w
\end{equation}
By calculating the low order process of $\mu^{-}$ decay and comparing the result with the Fermi theory, we can also get $G/\sqrt{2}=g^{2}/(8m^2_{w})$, from which we can decide the masses of $W^{\pm}$ particles. Then from Eq.(70), $Z^{0}$ particle's mass can also be determined. The result is completely the same as that in the current theory in which the Higgs mechanics is used. When mass eigen states are used, the gauge transformation of mass items is 
$$m^2_{w}{W}'^{+}_{\mu}{W}'^{-}_{\mu}+{1\over{2}}{m}^2_{Z}{Z'}_{\mu}{Z'}_{\mu}=m^2_{w}{W}^{+}_{\mu}{W}^{-}_{\mu}+{1\over{2}}{m}^2_{Z}{Z}_{\mu}{Z}_{\mu}$$
\begin{equation}
+{{\sin\vartheta_{w}}\over{2g}}(2\cos\vartheta_{w}{A}^3_{\mu}-2\sin\vartheta_{w}{B}_{\mu}+{{\sin\vartheta_{w}}\over{g}}\partial_{\mu}\theta)\partial_{\mu}\theta
\end{equation}
It is obviously variable under gauge transformation. In order to keep it unchanged, we can let
\begin{equation}
\partial_{\mu}\theta=-{{2g}\over\sin\vartheta_{w}}(\cos\vartheta_{w}{A}^3_{\mu}-\sin\vartheta_{w}{B}_{\mu})=-{{2g}\over{\sin\vartheta_w}}{Z}_{\mu}
\end{equation}
As shown before, the group parameter form of $U(1)$ field can be arbitrary. So in order to keep the mass items represented by mass eigen states unchanged under $SU(2)\times{U}(1)$ transformation, the form of group parameter $\theta$ can not be arbitrary. Eq.(72) should be satisfied. It is noted that according to the definition in Eq.(9), group parameter $\theta$ is finite. For infinitesimal transformation, we should let $\theta\rightarrow\theta\delta\lambda$ with $\partial_{\mu}\theta\delta\lambda=-2gZ_{\mu}\delta\lambda/\sin\theta_w$. In this way, the mass items of particles $W^{\pm}$ and $Z^{0}$ can be added into the Lagrangian directly without violating $SU(2)\times{U}(1)$ gauge invariability. Thus, when non-mass eigen states are used, we have
$$S_0=\int{d}^4{x}[-\bar{L}(\gamma_{\mu}\partial_{\mu}-ig{{\tau^{\alpha}}\over{2}}\gamma_{\mu}{A}^{\alpha}_{\mu}+ig'{1\over{2}}\gamma_{\mu}{B}_{\mu})L-\bar{l}_R(\gamma_{\mu}\partial_{\mu}+ig'\gamma_{\mu}{B}_{\mu}){l}_{R}$$
$$-{1\over{4}}(\partial_{\mu}{A}^{\alpha}_{\nu}-\partial_{\nu}{A}^{\alpha}_{\mu}+gf^{\alpha\beta\gamma}{A}^{\beta}_{\mu}{A}^{\gamma}_{\nu})^2-{1\over{4}}(\partial_{\mu}{B}_{\nu}-\partial_{\nu}{B}_{\mu})^2$$
$$-{1\over{2}}m^2_{A}{A}^{\alpha}_{\mu}{A}^{\alpha}_{\mu}-m_l(\bar{l}_{L}{l}_{R}+\bar{l}_{R}{l}_{L})-{1\over{2\zeta_{A}}}(\partial_{\mu}{A}^{\alpha}_{\nu})^2-{1\over{2\zeta_{B}}}(\partial_{\mu}{B}_{\mu})^2$$
$$+C^{+}_{\alpha}{R}_{\mu}\partial_{\mu}{C}_{\alpha}-gf^{\alpha\beta\gamma}{C}^{+}_{\alpha}{R}_{\mu}{A}^{\beta}_{\mu}{C}_{\gamma}+C^{+}\partial^{2}{C}+\bar{K}_1\delta\nu_{L}+\delta\bar{\nu}_{L}{K}_1$$
\begin{equation}
+\bar{K}_2\delta{l}_{L}+\delta\bar{l}_{L}{K}_2+\bar{K}_{3}\delta{l}_{R}+\delta\bar{l}_R{K}_3+\bar{K}_4\delta{B}_{\mu}+\delta\bar{B}_{\mu}{K}_4]
\end{equation}
According to Eqs. (9), (60) and (61), the infinitesimal transformations are
$$\delta\bar{\nu}_{L}={i\over{2}}(\theta_3-\theta)\bar{\nu}_{L}+i\theta_1\bar{l}_L~~~~~~~~~\delta\bar{l}_{L}={i\over{2}}(\theta_3+\theta)\bar{l}_{L}~~~~~~~~~\delta\bar{l}_{R}=i\theta\bar{l}_{R}$$ 
$$\delta\nu_{L}=-{i\over{2}}(\theta_3-\theta)\nu_{L}-i\theta_1{l}_L~~~~~~~~~\delta{l}_{L}=-{i\over{2}}(\theta_3+\theta){l}_{L}~~~~~~~~~\delta{l}_{R}=i\theta{l}_{R}$$
\begin{equation}
\delta{A}^{\alpha}_{\mu}=0~~~~~~\delta{B}_{\mu}=-{1\over{g}}\partial_{\mu}\theta~~~~~~\delta{C}^{+}_{\alpha}=\delta{C}_{\alpha}=\delta{C}=0~~~~~~~~\delta{C}^{+}=2\zeta_{B}\partial_{\mu}{B}_{\mu}
\end{equation}
Let $\theta_k=C_k\delta\lambda$, $\theta=C\delta\zeta$ similarly£¬ we have $(\delta\lambda)^2\rightarrow{0}$ so that $\delta^2\nu_{L}=\delta^2{l}_{L}=\delta^2{B}_{\mu}\rightarrow{0}$. By the same method shown before, renormalization can be done. If mass eigen states are used, by the transformation Eq.(63) and (64), we can also get the action which is also invariable under $SU(2)\times{U}(1)$ transformation. 
$$S_0=\int{d}^4{x}\{L_0(W^{\pm}_{\mu},Z_{\mu},A_{\mu},l,\nu)+i{g\over{\sqrt{2}}}[W^{+}_{\mu}\bar{\nu}\gamma_{\mu}(1+\gamma_5)l+W^{-}_{\mu}\bar{l}\gamma_{\mu}(1+\gamma_5)]$$
$$-i{{\sqrt{g^2+g'^2}}\over{4}}{Z}_{\mu}[\bar{\nu}\gamma_{\mu}(1+\gamma_5)\nu+\bar{l}\gamma_{\mu}(4\sin\theta_w-1-\gamma_5)l]-i{{gg'}\over{\sqrt{g^2+g'^2}}}A_{\mu}\bar{l}\gamma_{\mu}{l}$$
$$-m_l\bar{l}{l}-m^2_{w}{W}^{+}{W}^{-}-{1\over{2}}m^2_{Z}{Z}_{\mu}{Z}_{\mu}-{1\over{2\zeta_{A}}}[\partial_{\mu}{A}^{\alpha}_{\mu}(W^{\pm}_{\mu},{Z}_{\mu})]^2$$
$$-{1\over{2\zeta_B}}[\partial_{\mu}{B}_{\mu}(W^{\pm}_{\mu},Z_{\mu})]^2+C^{+}_{\alpha}{R}_{\mu}\partial_{\mu}{C}_{\alpha}-gf^{\alpha\beta\gamma}{C}^{+}_{\alpha}{R}_{\mu}{A}^{\beta}_{\mu}(W^{\pm}_{\mu},Z_{\mu}){C}_{\gamma}+C^{+}\partial^2{C}+\bar{K}_l\delta\nu$$
\begin{equation}
+\delta\bar{\nu}{K}_1+\bar{K}_2\delta{l}_L+\delta\bar{l}_L{K}_2+\bar{K}_3\delta{l}_R+\delta\bar{l}_{R}{K}_3+\bar{K}_4\delta{B}_{\mu}+\delta\bar{B}_{\mu}{K}_4\}
\end{equation}
In this case, the transformation rules of various fields becomes
$$\delta\nu=-{i\over{2}}(\theta_3-\theta)\nu-i\theta_1{l}~~~~~~~~~\delta\bar{\nu}={i\over{2}}(\theta_3-\theta)\nu+i\theta_1{l}~~~~~~~\delta{l}=-{i\over{2}}(\theta_3+\theta){l}$$
$$\delta\bar{l}={i\over{2}}(\theta_3+\theta){l}~~~~~~~~\delta{W}^{+}_{\mu}=0~~~~~~~\delta{W}^{-}_{\mu}=0~~~~~~~~\delta{Z}_{\mu}=-\sin\theta_{w}\delta{B}_{\mu}$$
\begin{equation}
\delta{A}_{\mu}={1\over{g}}\cos\theta_{w}\delta{B}_{\mu}~~~~~~~~~\delta{C}^{+}_{\alpha}=\delta{C}_{\alpha}=\delta{C}=0~~~~~~~~~~\delta{C}^{+}=\beta_{B}\partial_{\mu}{B}_{\mu}
\end{equation}
Let $\theta_k=C_{k}\delta\lambda$, $\theta=C\delta\lambda$ similarly, we have $(\delta\lambda)^2\rightarrow{0}$, $\delta^2\nu=\delta^2{l}=\delta^2{Z}_{\mu}=\delta^2{A}_{\mu}\rightarrow{0}$. The same renormalization can be carried out.
\par
In sum, because the Higgs particles can't be found up to now, it is still a big problem if there exists the Higgs particles. At present, some persons believe that the Higgs particles do not exist at all. Some theoretical modes have been put forward to replace the Higgs mechanics. For example, the Higgs particles are regarded as the bounding states of some new positive and anti-quark particles. But all these theories have some difficult problems. It can be said that the scheme provided in this paper is simplest and more rational without increasing any new particles or extra hypotheses. In order to ensure the invariability of motion equation of gauge fields under gauge transformation, this scheme is also necessary. The theory is meaningless if the invariability of motion equation of gauge fields can't be ensured. By giving up the principle of complete local gauge invariability and adopting the principle of incomplete local gauge invariability, we don't need the hypotheses of Higgs mechanics again. The description of theory can also become more simple and symmetrical.
\\
\\
References
\\
\\
(1) Dai Yuanbeng, Gauge Theory of Interaction, Science Publishing House, 23 (1987).\\
(2) Faddeev, V. N. Popov, Phys. Letters, 25B, 29 (1967). \\
(3)Hu Yaoguang, Gauge Theory of Field, Publishing House Of Huadong Normal University, 217£¨1984£©.\\
\end{document}